\documentclass[10pt,conference,a4paper]{IEEEtran}

\usepackage{amsmath}
\usepackage{graphicx}
\usepackage{bm}
\usepackage{cancel}
\usepackage{algorithm}
\usepackage{algorithmicx}
\usepackage{algpseudocode}
\usepackage{epstopdf}
\graphicspath{{../figures}{../figures/Experiment_Figures}}
\usepackage[font=footnotesize]{subfig}
\usepackage{pbox}
\usepackage{color}
\usepackage{setspace}
\usepackage{booktabs}
\usepackage{cite}
\usepackage{float}
\usepackage{multirow}
\usepackage{fullpage}
\usepackage{amssymb}

\usepackage{color}

\usepackage{amsthm}
\newtheorem{thm}{Theorem}

\def\Reals{\mbox{\rm I\kern-.2em R}} 
\ifCLASSINFOpdf
\else
\fi
\IEEEoverridecommandlockouts

\begin{document}
	%
	% paper title
	% Titles are generally capitalized except for words such as a, an, and, as,
	% at, but, by, for, in, nor, of, on, or, the, to and up, which are usually
	% not capitalized unless they are the first or last word of the title.
	% Linebreaks \\ can be used within to get better formatting as desired.
	% Do not put math or special symbols in the title.
	%\title{Improving convergence of data-driven basis for direct estimation of density functionals }
	\title{Direct Ensemble Estimation of Density Functionals }

	% author names and affiliations
	% use a multiple column layout for up to three different
	% affiliations
	\author{\IEEEauthorblockN{Alan Wisler\IEEEauthorrefmark{1},
			Kevin Moon\IEEEauthorrefmark{2}, Visar Berisha\IEEEauthorrefmark{1}
		}
		\IEEEauthorblockA{\IEEEauthorrefmark{1}Schools of ECEE and SHS, Arizona State University}
		\IEEEauthorblockA{\IEEEauthorrefmark{2}Genetics and Applied Math Departments, Yale University}
	}

	% make the title area
	\maketitle
	
	% As a general rule, do not put math, special symbols or citations
	% in the abstract
	\begin{abstract}
		Estimating density functionals of analog sources is an important problem in statistical signal processing and information theory. Traditionally, estimating these quantities requires either making parametric assumptions about the underlying distributions or using non-parametric density estimation followed by integration. In this paper we introduce a direct nonparametric approach which bypasses the need for density estimation by using the error rates of $k$-NN classifiers as ``data-driven" basis functions that can be combined to estimate a range of density functionals. However, this method is subject to a non-trivial bias that dramatically slows the rate of convergence in higher dimensions. To overcome this limitation, we develop an ensemble method for estimating the value of the basis function which, under some minor constraints on the smoothness of the underlying distributions, achieves the parametric rate of convergence regardless of data dimension. 
	\end{abstract}
	
	% no keywords

	\IEEEpeerreviewmaketitle

	\section{Introduction}
	%\label{sec:Intro}
	Density functionals that map probability density functions (PDFs) to $\Reals$ have been used in many signal processing applications involving classification \cite{moreno2003kullback}, segmentation \cite{hamza2003image}, source separation \cite{hild2001blind}, clustering \cite{banerjee2005clustering}, and other domains. Traditional estimation of these quantities typically relies on assuming a parametric model for the underlying PDFs, and calculating the desired functional from the estimated parameters of that model. Parametric methods offer good mean squared error (MSE) convergence rates $\mathcal{O}(N^{-1})$ ($N$ represents the number of samples) when an accurate parametric model is known, but become asymptotically biased if the data do not fit the assumed model. To guarantee an asymptotically consistent estimator in these scenarios, two general classes of estimators exist: 1) non-parametric plug-in estimators and 2) graph-based direct estimators \cite{hero01}.

	Non-parametric plug-in estimators have been used for estimating functionals of both discrete and analog distributions \cite{wu2016minimax, jiao2015minimax, wang2009universal}, with a particular focus on entropy estimation \cite{valiant2010clt, valiant2011estimating}. While non-parametric plug-in estimators don't require a parametric model, they generally have high variance, are sensitive to outliers, and scale poorly with dimension \cite{hero01}. Alternatively, graph-based estimators, exploit the asymptotic properties of minimal graphs to \textit{directly} estimate density functionals without estimating the underlying distributions. These methods have been used to estimate density functionals such as entropy \cite{hero1998robust}, the $\alpha$-divergence \cite{hero01}, and the $D_p$-divergence \cite{berisha2014empirically}. Graph based methods bypass the complication of fine tuning parameters such as kernel bandwidth or histogram bin size and can offer faster convergence rates in some scenarios \cite{hero01}. In this paper, we attempt to overcome two of the fundamental limitations of previously derived graph-based estimators: their specificity (their inability to estimate a broad variety of density functionals) and their convergence rates in higher dimensional spaces.
	
	The first major limitation of graph-based estimators is their specificity to the density functional being estimated. Whereas plug-in estimation methods can use the same general procedure to estimate a wide range of density functionals, most graph-based estimators can only be used to estimate a specific density functional. To overcome this, we use a set of density functionals, which can be estimated by a corresponding set of graph-based estimators, as data-driven basis functions. We can then use linear combinations of these data-driven basis functions to estimate unknown density functionals that cannot be estimated via graph-based methods directly. This strategy was originally introduced in \cite{wisler2017data} using a new basis set similar to the deterministic Bernstein polynomial basis. This paper extends that work by re-interpreting the $k$-NN classifier error rate as a data-driven basis function. 
	
	In contrast to the Bernstein-like basis in \cite{wisler2017data}, much is already known about the finite-sample convergence properties of the $k$-NN error rate. The $k$-NN error rate converges in the MSE to its asymptotic value at the parametric rate when $d\leq 4$, but slows to $\mathcal{O}(N^{-\frac{4}{d}})$ at higher dimensions \cite{psaltis1994finite}. To overcome the slow convergence rate, we generalize theory previously developed for ensemble estimation of density functionals \cite{sricharan2013ensemble,moon2014multivariate,moon2016improving} to develop ensemble estimators of the $k$-NN error rate that can guarantee a $\mathcal{O}(N^{-1})$ rate of convergence independent of dimension. In  \cite{sricharan2013ensemble,moon2014multivariate,moon2016improving}, ensembles were used in the context of density estimation by varying the bandwidth parameter. We generalize this approach through an ensemble formed by {\em varying the sample size} instead; we show that this approach can yield the same parametric convergence rate. By improving the convergence rate in our estimates of the basis functions, we also improve the convergence rate of the resulting density functional estimators. 
	\vspace{-.09cm}
	\section{Graph-Theoretic Basis Functions}
	\vspace{-.09cm}
	
	Consider the problem in which we are given a set of data $[\mathbf{X},\mathbf{y}]$ containing $N$ instances, where each instance is represented by a $d$-dimensional feature vector $\mathbf{x}_i$ and a binary label $y_i$. Suppose that this data is sampled from an unknown underlying distribution, $f_\mathbf{x}(\mathbf{x})$, where $f_\mathbf{x}(\mathbf{x})=p_0f_0(\mathbf{x})+p_1f_1(\mathbf{x})$ consists of the two conditional class distributions $f_0(\mathbf{x})$ and $f_1(\mathbf{x})$ for classes $0$ and $1$, with prior probabilities $p_0$ and $p_1$ respectively. The posterior likelihood of class 1, $\eta(\mathbf{x})$, evaluated at a point $\mathbf{x}=\mathbf{x}^*$, is
	\begin{equation}\eta(\mathbf{x}^*)=P[y=1|\mathbf{x}=\mathbf{x}^*]=\frac{p_1f_1(\mathbf{x}^*)}{f_\mathbf{x}(\mathbf{x}^*)}.
	\label{eq:posterior}
	\end{equation} 
	Many density functionals, including all $f$-divergences, can be expressed in terms of $\eta(\mathbf{x})$ as 
	\begin{equation}
	G(f_0,f_1)=\mathbb{E}_f[g(\eta)]=\int g(\eta)f_{\mathbf{x}}(\mathbf{x}) d\mathbf{x},
	\label{eq:form}
	\end{equation}
	where $g(\eta)$ is some function corresponding to the functional $G(f_0,f_1)$. Additionally, Cover and Hart \cite{cover1967nearest} showed that the error rate of a $k$-NN classifier trained on $N$ samples converges to
	\begin{equation}
	\begin{aligned}
	\lim\limits_{N\rightarrow \infty}R_k(N)=R_k(\infty)=E_f[r_k(\eta)],
	\end{aligned}
	\end{equation}
	where                                                
	\begingroup\makeatletter\def\f@size{8}\check@mathfonts
	\def\maketag@@@#1{\hbox{\m@th\normalsize\normalfont#1}}%
	\begin{equation}
	r_k(\eta)=\sum_{i=\lceil \frac{k}{2}\rceil}^{k}\binom{k}{i}\Big(\eta^i(1-\eta)^{k-i+1}+(1-\eta)^i\eta^{k-i+1}\Big).
	\label{eq:r_k}
	\end{equation}
	\endgroup
	Rather than use these error rates to measure the performance of a $k$-NN classifier, as is most common, we exploit our knowledge of the asymptotic properties of these error rates to estimate other desirable quantities. We thus re-interpret the error rates of $k$-NN classifiers for a range of $k$ values $k\in\mathcal{K}=[1,...,K]$ as basis functions. Suppose that we wish to estimate some $G(f_0,f_1)$ in a regime for which traditional plug-in estimators under-perform. In our previous work \cite{wisler2017data}, we identified a fitting criterion for identifying weights $\boldsymbol{\alpha}=\alpha_1,...,\alpha_K$ that minimize the residual 
	\begin{equation}
	\left | g(\eta) -  \sum_{k\in\mathcal{K}}^{}\alpha_k r_k(\eta) \right |^2,
	\label{eq:g_approx}
	\end{equation}
	when both the basis functions and the density functional can be expressed in terms of the posterior (as in Eqns. \eqref{eq:form} and \eqref{eq:r_k}). If we use the same approach to learn the optimal weights, then, for sufficient $N$, we approximate the density functional by
	\begin{equation}
	G(f_0,f_1)\approx \sum_{k\in\mathcal{K}}^{}\alpha_k \mathbb{E}_f[r_k(\eta)] =\hat{G}(\mathbf{X},\boldsymbol{\alpha})
	\label{eq:G_estimator1}
	\end{equation}
	This estimator $\hat{G}(\mathbf{X},\boldsymbol{\alpha})$ makes no assumptions on the underlying distributions and bypasses the need for density estimation. However, finite sample estimates of the $k$-NN error rate contain a non-trivial bias rate which leads to glacially slow convergence at higher dimensions. To overcome this bias, we use optimally weighted ensemble estimators to achieve fast MSE convergence regardless of the dimension.
	\vspace{-.1cm}
	\section{Ensemble Estimation of the Basis}
	\vspace{-.1cm}
	\label{sec:ensemble_method}
	
	If we assume that densities $f_0(\mathbf{x})$ and $f_1(\mathbf{x})$  1) are absolutely continuous over $\Reals^n$, 2) have compact support, 3) have $s$ continuous derivatives, and 4) vanish close to the boundaries; then the bias of the  $k$-NN error rate from a classifier trained on $N$ samples is
	\begin{equation}
	\mathbb{B}[R_k(N)]=\sum_{j=2}^{s-1}c_j N^{-\frac{j}{d}}+\mathcal{O}(N^{-\frac{s}{d}}),
	\label{eq:bias}
	\end{equation}
	where $d$ represents the intrinsic dimension of the distribution and the expansion constants $c_j$ depend generally upon $k$, the Euclidean metric used to generate the $k$-NN graph, and the underlying distributions \cite{psaltis1994finite}. Asymptotically the first term in the sum will dominate, and the bias of $R_k(N)$ approaches zero at a rate of $\mathcal{O}(N^{-\frac{2}{d}})$. For low dimensional problems ($d\leq 4$) the squared bias matches the MSE parametric rate $\mathcal{O}(N^{-1})$. However the bias converges glacially slow at higher dimensions.
	
	To improve the rate of convergence, we use an ensemble of finite sample estimators to cancel the lower order bias terms. Suppose that we have $L$ $k$-NN classifiers, each trained on $M_i=l_iN$ samples, where $l_i \leq 1$ and $i \in \mathcal{I}= [1, \dots, L]$. Additionally, assume that each subsample is evaluated on $N$ held-out data points.  The ensemble estimate of the error rate is
	\begin{equation}
	\Phi_k(N,\mathbf{w}^*)=\sum_{i\in\mathcal{I}}^{}w_i R_k(N,l_i)
	\end{equation}
	where $\mathbf{w}^*=[w_1,w_2,...,w_L]$.	If $\sum_{i\in\mathcal{I}}^{}w_i=1$, $\Phi_k(N,\mathbf{w}^*)$ is
	\begingroup\makeatletter\def\f@size{9}\check@mathfonts
	\def\maketag@@@#1{\hbox{\m@th\normalsize\normalfont#1}}%
	\begin{equation}
	\begin{aligned}
	\mathbb{B}[\Phi_k(N,&\mathbf{w}^*)]=\mathbb{E}\bigg[\sum_{i\in\mathcal{I}}^{}w_i R_k(N,l_i)\bigg]-R_k(\infty)\\
	&=\sum_{i\in\mathcal{I}}^{}w_i\bigg[\sum_{j=2}^{s-1}c_j M_i^{-\frac{j}{d}}+\mathcal{O}(N^{-\frac{s}{d}})\bigg].
	\end{aligned}
	\label{eq:ensemble_bias1}
	\end{equation}
	\endgroup
	%	We can separate the bias terms in \eqref{eq:ensemble_bias1} into two groups
	%
	%	\begin{equation}
	%	\mathbb{B}[\Phi_k(N,\mathbf{w}^*)]=\sum_{i\in\mathcal{I}}^{}w_i\Bigg[\sum_{j=2}^{\lceil \frac{d}{2}-1\rceil}c_j M_i^{-\frac{j}{d}}+\sum_{j=\lceil \frac{d}{2}\rceil}^{\infty}c_j M_i^{-\frac{j}{d}}\Bigg].
	%	\label{eq:ensemble_bias2}
	%	\end{equation} 
	If $s\geq \lceil \frac{d}{2}\rceil$, we can ensure that the bias convergence rate is of order $\mathcal{O}(N^{-\frac{1}{2}})$ by selecting weights which set all terms of $j<\lceil \frac{d}{2}\rceil$ to zero: 
	\begin{equation}
	\begin{aligned}
	w_1,...,w_L=\underset{w_1,...,w_L}{\mathrm{argmin}} &\quad \sum_{i\in\mathcal{I}}^{}w_i^2\\
	\text{Subject to } & \sum_{i\in\mathcal{I}}^{}w_i=1\\
	&	\sum_{i\in\mathcal{I}}^{} w_i l_i^{-\frac{j}{d}}=0 , j\in \mathcal{J}
	\end{aligned}
	\label{eq:Method1}
	\end{equation} 
	where $\mathcal{J}= [2,...,\lceil d/2-1 \rceil]$. Note that as long as $L\geq \lceil d/2-1 \rceil$ and all $l_i$ are assigned unique values, \eqref{eq:Method1} is guaranteed to have at least one solution\cite{sricharan2013ensemble}. Additionally, since the variance of each of the subsample estimators converges at a rate of $\mathcal{O}(N^{-1})$ the variance of a linear combination of these estimators will converge at $\mathcal{O}(N^{-1})$\cite{sricharan2013ensemble}. Thus the MSE of $\Phi_k(N,\mathbf{w}_0)$ will converge at rate  $\mathcal{O}(N^{-1})$.
	
	This optimization criterion imposes no constraints on the magnitude of the weights and empirically leads to excessively high variance estimates, making it impractical for small $N$ \cite{sricharan2013ensemble,moon2014multivariate,moon2016improving}. A suggested solution in \cite{sricharan2013ensemble,moon2014multivariate,moon2016improving} is to relax the optimization criteria so that, rather than setting the bias terms to zero, they are bounded by a bias threshold $\epsilon_1$ scaled by $N^{-\frac{1}{2}}$, which is minimized subject to a fixed variance threshold which we will call $\epsilon_2$. However, when $\epsilon_2 \leq ||w_0||_2^2$, the bias threshold $\epsilon_1$ can become functionally dependent on $N$, which could potentially slow the rate of convergence. As an alternative, we propose setting the bias and variance thresholds equal to each other ($\epsilon_1=\epsilon_2=\epsilon$), thus allowing the variance threshold to scale with the bias threshold. This ensures that we maintain the desired rate of convergence asymptotically while still controlling the variance at lower $N$. The resulting fitting routine is
	\begin{equation}
	\allowdisplaybreaks
	\begin{aligned}
	\mathbf{w}_r=w_1,...&,w_L=\underset{w_1,...,w_L}{\mathrm{argmin}} \quad \epsilon \\
	\text{Subject to } & \sum_{i\in\mathcal{I}}^{}w_i=1\\
	\sum_{i\in\mathcal{I}}^{} &w_i l_i^{-\frac{j}{d}}\leq \epsilon N^{\frac{j}{d}-\frac{1}{2}}\quad,j\in \mathcal{J} \\
	& \sum_{i\in\mathcal{I}}^{} w_i^2 \leq \lambda \epsilon,
	\end{aligned}
	\label{eq:Method_relaxed2}
	\end{equation} 
	where $\lambda$ is a tuning parameter used to control the trade-off between minimizing the bias and the variance. In theorem 1 we show that the MSE convergence rate of this new estimator is of order $\mathcal{O}(N^{-1})$.
	
	\begin{thm}
		\label{thm:Main}
		If there exists a set of weights $\mathbf{w}_0$ satisfying the constraints of \eqref{eq:Method1}, then 
		
		\begin{equation}
		\mathbb{E}\Big[\big(\Phi_k(N,\mathbf{w}_r)-R_k(\infty)\big)^2\Big]= \mathcal{O}(N^{-1}).
		\end{equation}
	\end{thm}
	
	\textbf{Proof:}
	\textit{Lemma 1:}	If there exists a set of weights $\mathbf{w}_0$ satisfying the constraints of \eqref{eq:Method1}, then $\epsilon$ in \eqref{eq:Method_relaxed2} is bounded by 
	\begin{equation*}
	\epsilon\leq \epsilon^*=||\mathbf{w}_0||_2^2
	\end{equation*}
	\textit{Proof of Lemma 1:} Suppose that $\epsilon> \epsilon^*$. Since $(\epsilon^*,\mathbf{w}_0)$ satisfies the constraints of \eqref{eq:Method_relaxed2}, $\epsilon^*$ violates the minimality of $\epsilon$. Therefore by contradiction $\epsilon\leq \epsilon^*$.
	
	Now	we can define the bias of $\Phi_k(N,\mathbf{w}_r)$ as
	\begingroup\makeatletter\def\f@size{9}\check@mathfonts
	\def\maketag@@@#1{\hbox{\m@th\normalsize\normalfont#1}}%
	\begin{equation}
	\begin{aligned}
	\mathbb{B}[\Phi_k(N,\mathbf{w}_r)]&=\sum_{j=2}^{\lceil \frac{d}{2}-1\rceil}c_j\sum_{i\in\mathcal{I}}^{}w_i M_i^{-\frac{j}{d}}+\mathcal{O}(N^{-\frac{1}{2}})\\
	&\leq \sum_{j=2}^{\lceil \frac{d}{2}-1\rceil} c_j \epsilon N^{-\frac{1}{2}}+\mathcal{O}(N^{-\frac{1}{2}})
	\end{aligned}
	\end{equation}
	\endgroup
	Using Lemma 1, this can be upper bounded by 
	\begin{equation}
	\mathbb{B}[\Phi_k(N,\mathbf{w}_r)]\leq \sum_{j=2}^{\lceil \frac{d}{2}-1\rceil} c_j \epsilon^* N^{-\frac{1}{2}}+\mathcal{O}(N^{-\frac{1}{2}})
	\end{equation}
	thus ensuring that $\mathbb{B}[\Phi_k(N,\mathbf{w}_r)]=\mathcal{O}(N^{-\frac{1}{2}})$. Similarly, we can express the variance as
	\begin{equation}
	\begin{aligned}
	\mathbb{V}[\Phi_k(N,\mathbf{w}_r)]\leq \frac{\epsilon}{4N}\leq \frac{\epsilon^*}{4N}=\mathcal{O}(N^{-1}).
	\end{aligned}
	\end{equation}
	Since both $\mathbb{B}[\Phi_k(N,\mathbf{w}_r)]^2$ and $\mathbb{V}[\Phi_k(N,\mathbf{w}_r)]$ are bounded by rate $\mathcal{O}(N^{-1})$, the MSE must be bound by this rate as well.
	
	It is worth noting that while Theorem \ref{thm:Main} guarantees $\mathcal{O}(N^{-1})$ convergence asymptotically, it may converge more slowly for specific ranges of $N$.
	
	Returning to the original problem of estimating an unknown density functional $G(f_0,f_1)$, we can update the estimator $\hat{G}(\mathbf{X},\boldsymbol{\alpha})$ described in \eqref{eq:G_estimator1} to be a linear combination of these ensemble estimators
	\begin{equation}
	\hat{G}(\mathbf{X},\boldsymbol{\alpha})=\sum_{k\in\mathcal{K}}^{}\alpha_k \Phi_k(N,\mathbf{w}_r).
	\label{eq:Combined_Estimator}
	\end{equation}
	%	We postulate that if all of the basis functions achieve 
	If $\hat{G}(\mathbf{X},\boldsymbol{\alpha})$ is an asymptotically consistent estimate of $G(f_0,f_1)$, then the bias is
	%	\begin{equation}
	%	\lim\limits_{N\rightarrow \infty}\hat{G}(\mathbf{X},\boldsymbol{\alpha})=G(f_0,f_1)
	%	\end{equation}
	%	the bias and variance of this estimator are
	\begin{equation}
	\mathbb{B}[\hat{G}]=\sum_{k\in\mathcal{K}}^{}\alpha_k 	\mathbb{B}[\Phi_k(N,\mathbf{w}_r)],
	\label{eq:bias_combined}
	\end{equation}
	and we can bound the variance by
	\begingroup\makeatletter\def\f@size{8.5}\check@mathfonts
	\def\maketag@@@#1{\hbox{\m@th\normalsize\normalfont#1}}%
	\begin{equation}
	\mathbb{V}[\hat{G}]\leq\sum_{k\in\mathcal{K}}^{}\sum_{j\in\mathcal{K}}^{}\alpha_j \alpha_k 	\sqrt{\mathbb{V}[\Phi_j(N,\mathbf{w}_r)]\mathbb{V}[\Phi_k(N,\mathbf{w}_r)]}.
	\label{eq:var_combined}
	\end{equation}
	\endgroup
	Since all of the basis functions comprising $\hat{G}(\mathbf{X},\boldsymbol{\alpha})$ achieve an MSE convergence rate of $\mathcal{O}(N^{-1})$, the combined estimator also converges at rate $\mathcal{O}(N^{-1})$.

	\section{Numerical Results}

	\label{sec:simulation_overview}
	To evaluate the efficacy of the method proposed in this paper we consider the problem of estimating the Hellinger distance between two multivariate normal distributions. In this experiment each class PDF is distributed according to  $f(\mathbf{x})\sim N(\mu \mathbf{1}_d,\mathbf{\Sigma}_d)$, where the $i,j$ term in $\mathbf{\Sigma}_d$ is defined by $\sigma_{i,j}=\beta^{|i-j|}$ for all $i,j\in [1,\dots,d]$. We set $\mu=0$ and $\beta=0.8$ for distribution $f_0(\mathbf{x})$ and $\mu=1$ and $\beta=0.9$ for distributions $f_1(\mathbf{x})$. This set of parameters offers each distribution a unique elliptical covariance structure and was identified as particularly challenging for direct estimation in \cite{wisler2017data}. 
	%	\begin{table}[]
	%		\centering
	%		\caption{Experiment overview table}
	%		\label{tab:experiment_outline}
	%		\begin{tabular}{l|l|l|l|l|}
	%			\cline{2-5}
	%			& \multicolumn{2}{l|}{$f_0(\mathbf{x})$} & \multicolumn{2}{l|}{$f_1(\mathbf{x})$}   \\ \cline{2-5} 
	%			      & $\mu$      & $\beta$       & $\mu$                & $\beta$ \\ \hline
	%			\multicolumn{1}{|l|}{Experiment 1}       & 0          & 0            & $\sqrt{\frac{1}{d}}$ & 0       \\
	%			\multicolumn{1}{|l|}{Experiment 2}       & 0          & 0.8          & $1$ & 0.8     \\
	%			\multicolumn{1}{|l|}{Experiment 3}      & 0          & 0.8          & $\sqrt{\frac{1}{d}}$ & 0.9       \\ \hline
	%		\end{tabular}
	%	\end{table}
	%	
	%	Each of these simulations utilize the following parameters:
	%	\begin{itemize}
	%		\item Monte Carlo iterations : 1000
	%		\item $N=[100,193,373,720,1389,2683,5179,10000]$
	%		\item $\mathbf{l}=[5,6.16,7.6,9.37,11.55,14.24,17.56,21.64,\\26.68,32.90,40.56,50]\times 10^{-2}$
	%		\item $\mathbf{k}=[1,3,5,7,9]$
	%		\item $d$=10
	%	\end{itemize}
	
		\begin{figure*}[!htb]
			\begin{center}
				\subfloat[Estimates of $R_1$]{
					\includegraphics[width=0.49\textwidth]{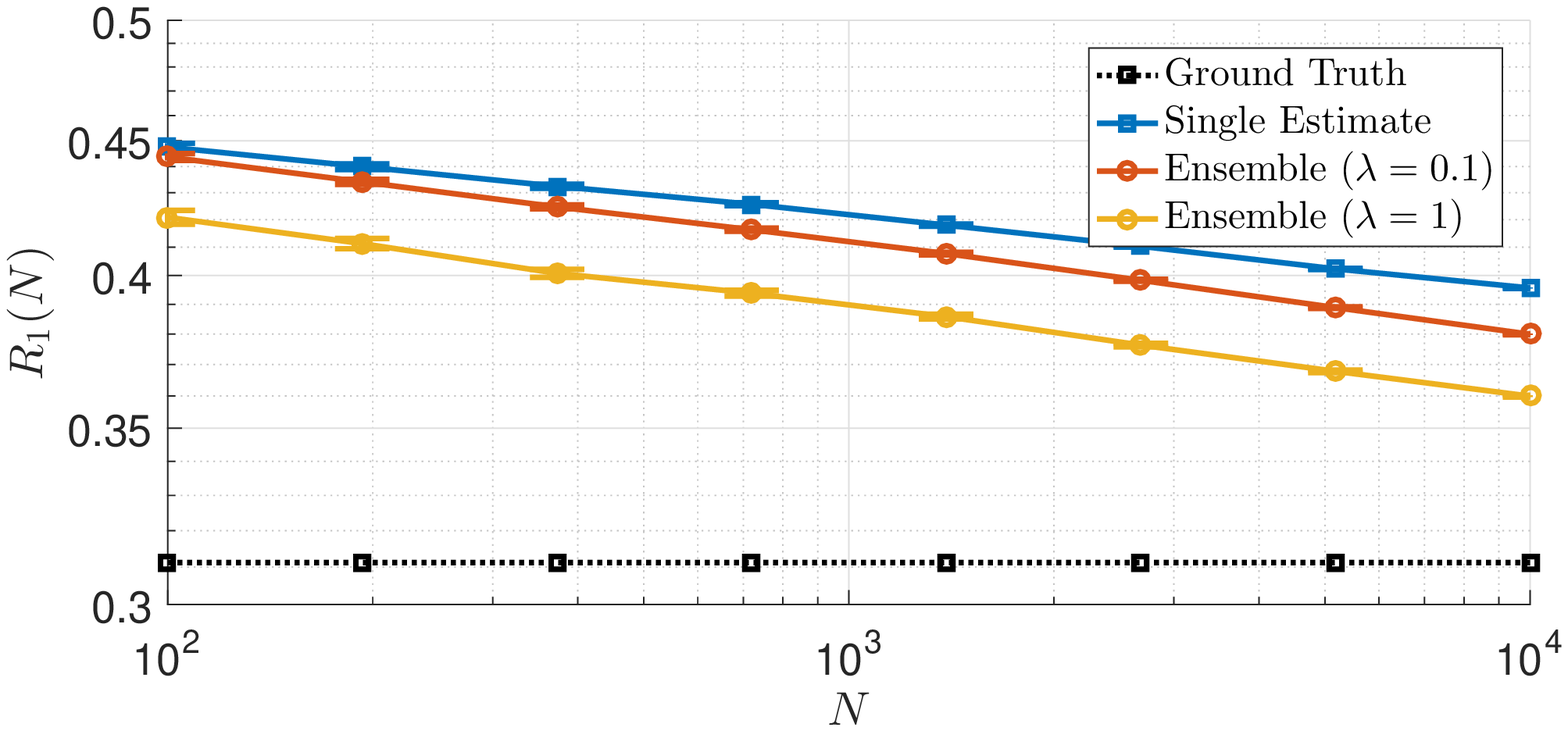}
					\label{fig:k1_dplot}
				}	
				\subfloat[MSE of $R_1$ estimates]{
					\includegraphics[width=0.49\textwidth]{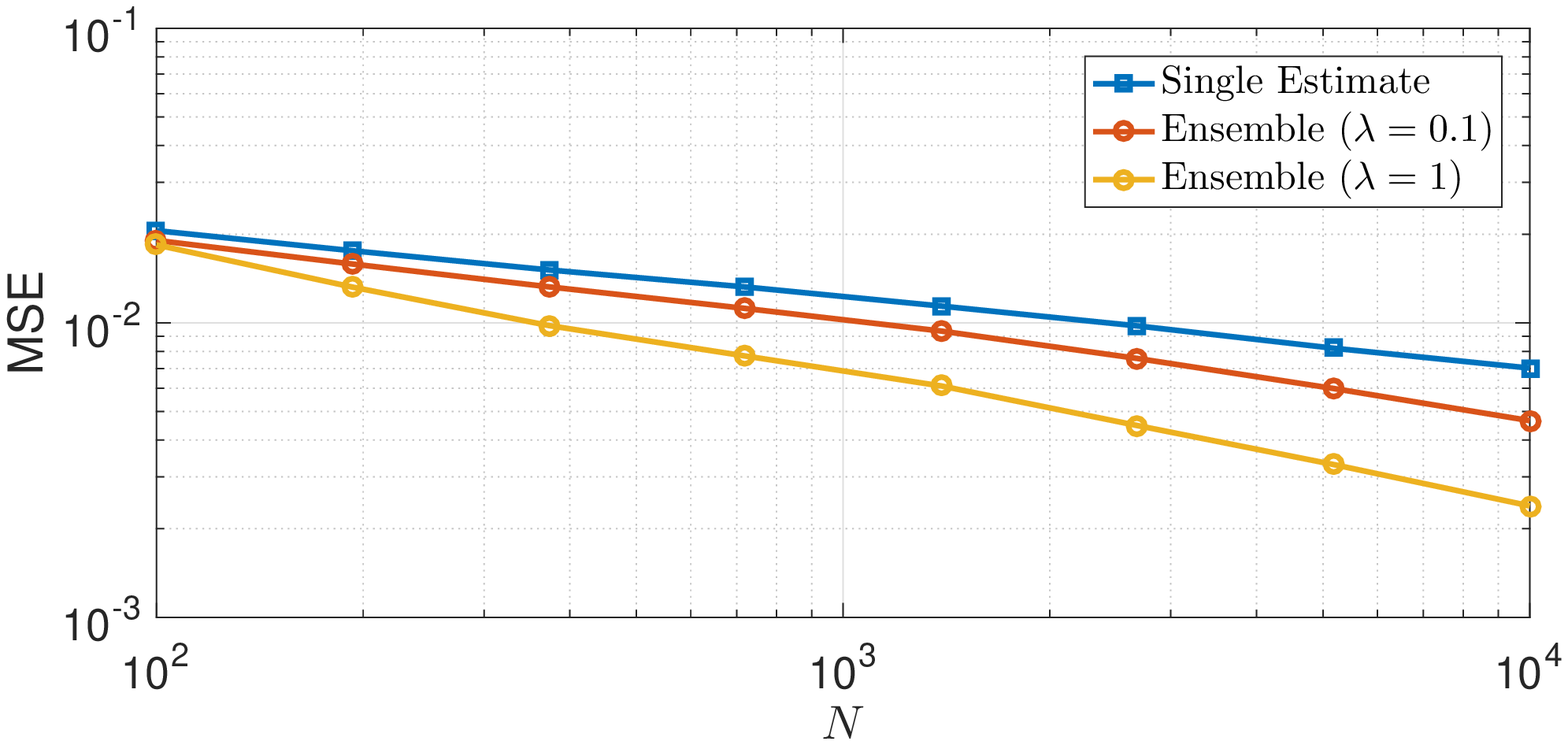}
					\label{fig:k1_MSE}
				}\\
				\subfloat[Estimates of $H^2(f_0,f_1)$]{		
					\includegraphics[width=0.49\textwidth]{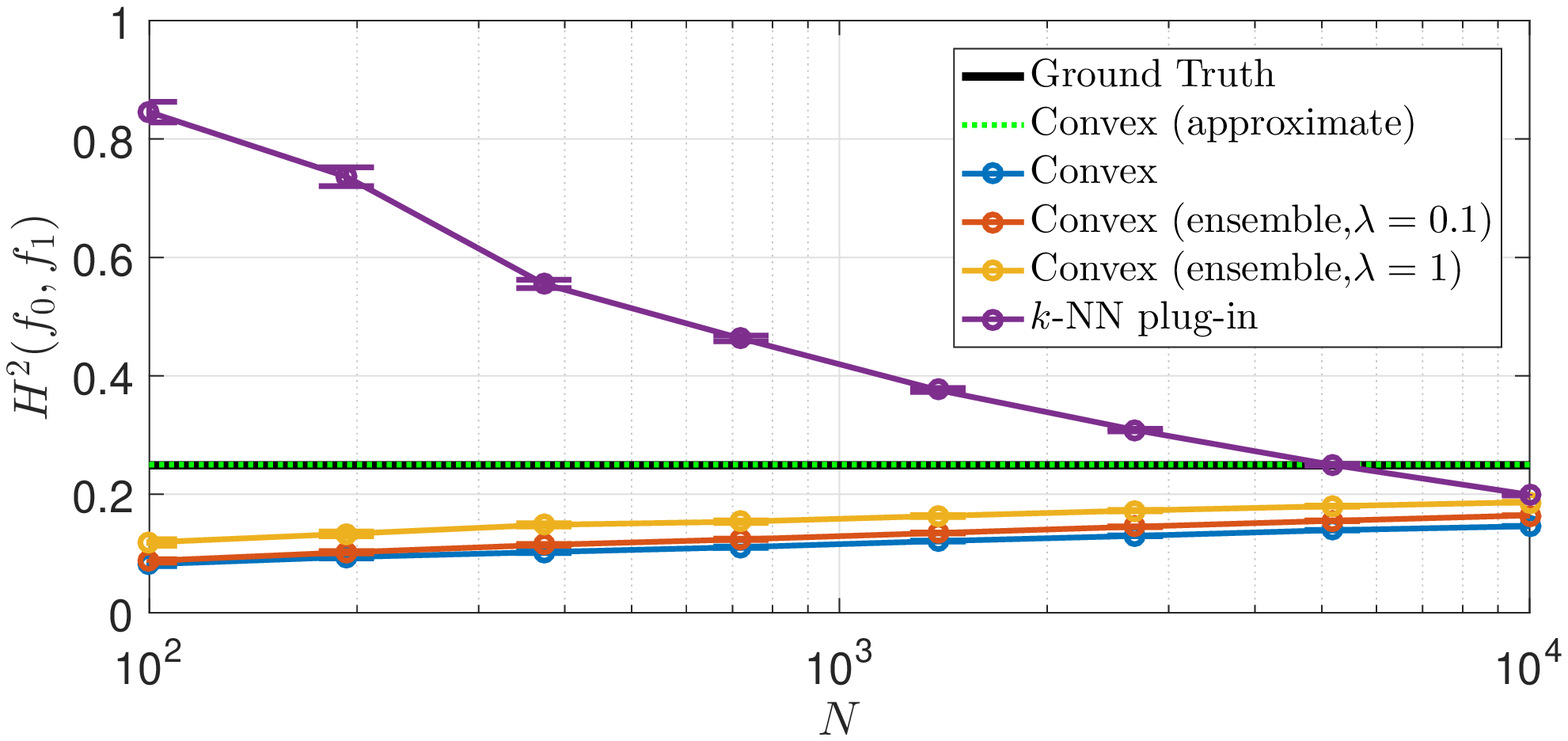}
					\label{fig:Hellinger_dplot}
				}
				\subfloat[MSE of  $H^2(f_0,f_1)$ estimates]{
					\includegraphics[width=0.49\textwidth]{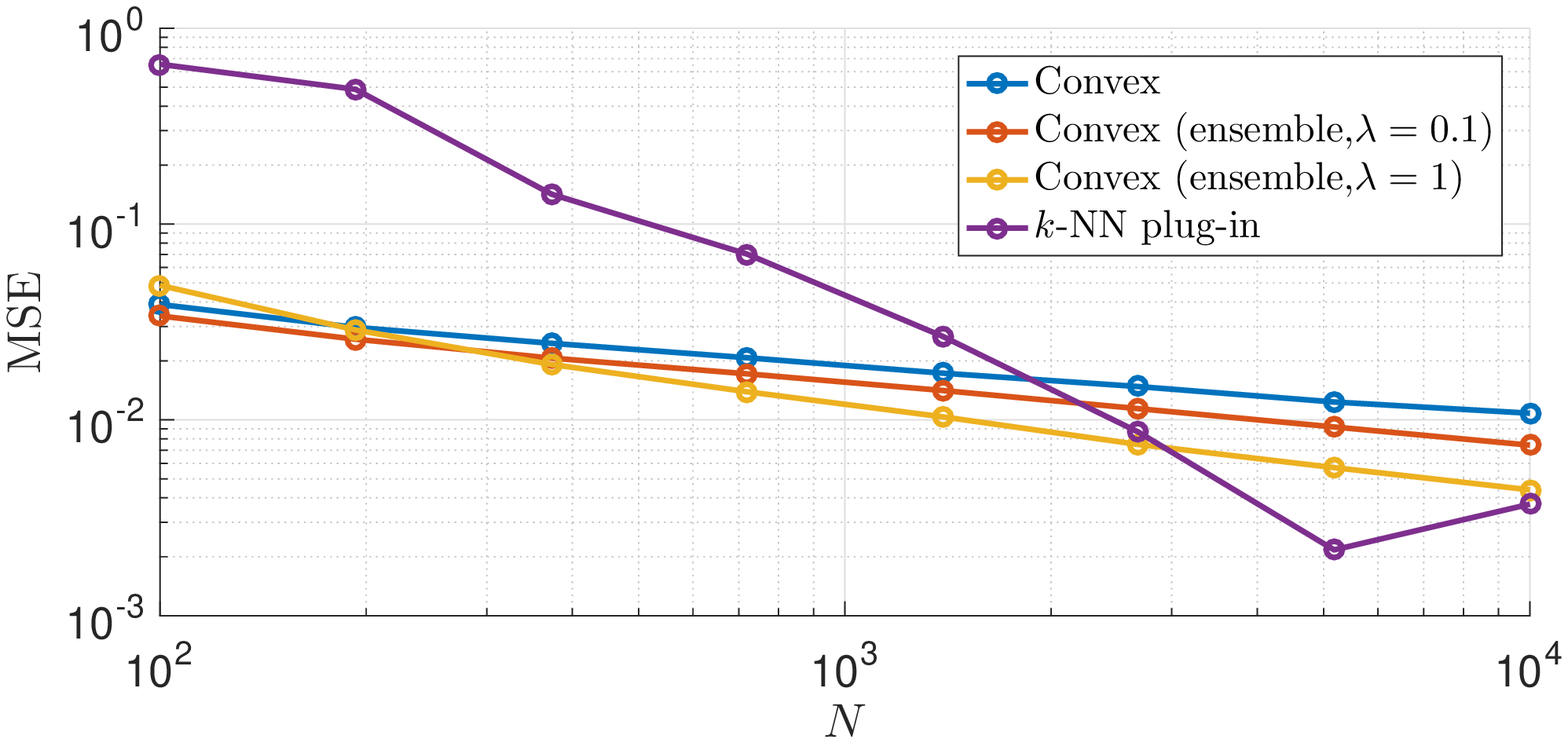}
					\label{fig:Hellinger_MSE}
				}	 				
				\caption{(a) \& (c) display the mean and standard error across sample size for each estimator of the $1$-NN classifier error rate and Hellinger-squared distance respectively, while (b) \& (d) display the corresponding MSE results for these estimators. These results reflect the average of $1000$ Monte Carlo trials where the $12$ $l_i$ values are logarithmically spaced between $0.05$ and $0.5$ and $\mathbf{k}=[1,3,5,7,9]$.}
				\label{fig:Exp3Plots}
			\end{center}
		\end{figure*}	
	
	At each sample size $N$, we compare 3 methods of estimating $R_k$ for all $k\in \mathbf{k}$. The first method is to simply evaluate the error rate of a $k$-NN classifier on the held-out portion of the data. We then generate ensemble estimates using the relaxed method described in Section \ref{sec:ensemble_method} using $\lambda=0.1$ and $\lambda=1$. For the $k=1$ basis, the resulting estimates are displayed relative to the ground truth in Figure \ref{fig:k1_dplot}, and the corresponding MSE is displayed in Figure \ref{fig:k1_MSE}. We see that while all three estimates contain some finite-sample bias, the bias is significantly reduced using the proposed method, particularly when $\lambda$ is set to prioritize minimization of the bias more heavily. As a result of the reduced bias, we see a significant improvement in the rate at which the MSE decays with $N$ in Figure \ref{fig:k1_MSE}.	It is worth noting that, despite the improvement, the slope of the MSE for the ensemble methods here is not indicative of the $\mathcal{O}(N^{-1})$ rate guaranteed by the proposed ensemble method since the theorem does not guarantee that it converges consistently at this rate across all $N$. As we increase the sample size, we expect that the rate would eventually reach the expected $\mathcal{O}(N^{-1})$.
	
	Next we evaluate the ensemble method when using $R_k(N)$ as a basis set for estimating density functionals like the squared Hellinger distance:
	\begin{align}
	H^2(f_0, f_1) & = \frac{1}{2} \int \left ( \sqrt{f_0(\mathbf{x})} - \sqrt{f_1(\mathbf{x})} \right)^2 d\mathbf{x} \\
	& = \frac{1}{2} \int g(\eta) f_{\mathbf{x}} (\mathbf{x}) d\mathbf{x},
	\end{align}	
	where $g(\eta) = (\sqrt{\eta} - \sqrt{1-\eta})^2$. Estimates of the Hellinger distance are generated using the optimal fitting weights, $\boldsymbol{\alpha}$, for estimating $g(\eta)$ (using the fitting criterion in \cite{wisler2017data}) and using the ensemble weights \eqref{eq:Method_relaxed2} to combine each of the 3 previously described estimators for $R_k(N)$ into an estimate of $H^2(f_0,f_1)$. Since the basis weights, $\boldsymbol{\alpha}$, are estimated using the convex optimization procedure outlined in \cite{wisler2017data}, we refer to this as the convex method. In addition to the three estimates acquired in this manner, we compare against a non-parametric density ($k$-NN) estimation+plug-in strategy for estimating the Hellinger distance which is calculated using the universal divergence estimation approach described in \cite{sutherland2012kernels} and implemented in the ITE toolbox \cite{szabo14information}.

	Figure \ref{fig:Hellinger_dplot} displays the predicted values of each method relative to the ground truth. These results confirm that the bias reduction achieved in the basis set translates to a similar improvement in the combined estimator.  In this experiment, the plug-in estimator exhibits a much larger bias at lower $N$, but appears to achieve faster convergence. Observing this estimator at higher $N$, shows that it crosses the true value and is significantly biased. In the MSE results, we see that at lower $N$ selecting the smaller $\lambda$ value yields better performance, while the higher $\lambda$ performs best for larger $N$. This matches our expectation that bias reduction becomes a higher priority as $N$ increases, and a corresponding increase in $\lambda$ is appropriate.

	%	%	% 
	\section{Conclusion}
	In this paper we consider using a weighted combination of $k$-NN error rates in order to estimate unknown density functionals. To improve the convergence rate of this approach, which slows dramatically at higher dimensions, we develop an ensemble estimate of the $k$-NN error rate which can guarantee $\mathcal{O}(N^{-1})$ convergence regardless of dimension, when the densities are sufficiently smooth. We evaluate the efficacy of this approach by estimating the Hellinger distance for a pair of multivariate Gaussian distributions on 10-dimensional data. In this scenario, our approach generally outperformed a plug-in estimator that first requires non-parametric density estimation. 
	
		\section{Acknowledgements}
		 The authors gratefully acknowledge Dennis Wei and Karthikeyan Ramamurthy for their help in discussing the ideas presented in this paper. This research was supported in part by Office of Naval Research grant N000141410722 (PI: Berisha).

	%		\begin{thebibliography}{1}
	%			
	%			\bibitem{IEEEhowto:kopka}
	%			H.~Kopka and P.~W. Daly, \emph{A Guide to \LaTeX}, 3rd~ed.\hskip 1em plus
	%			0.5em minus 0.4em\relax Harlow, England: Addison-Wesley, 1999.
	%			
	%		\end{thebibliography}
	\bibliographystyle{IEEEtran}
	%	 argument is your BibTeX string definitions and bibliography database(s)
	\bibliography{References,ReferencesVB}
\end{document}